# Phonon Hydrodynamic Heat Conduction and Knudsen Minimum in Graphite

Zhiwei Ding[1, †], Jiawei Zhou[1, †], Bai Song[1], Vazrik Chiloyan[1], Mingda Li[1, 2], Te-Huan Liu[1], and Gang Chen[1,*]

[1] Department of Mechanical Engineering, Massachusetts Institute of Technology, Cambridge, MA, 02139, USA

[2] Department of Nuclear Science and Engineering, Massachusetts Institute of Technology, Cambridge, MA, 02139, USA

[†]These authors contributed equally to this work.

[*]Author to whom correspondence should be addressed. E-mail: gchen2@mit.edu

**Abstract**

In the hydrodynamic regime, phonons drift with a nonzero collective velocity under a temperature gradient, reminiscent of viscous gas and fluid flow. The study of hydrodynamic phonon transport has spanned over half a century but has been mostly limited to cryogenic temperatures (~1 K) and more recently to low-dimensional materials. Here, we identify graphite as a three-dimensional material that supports phonon hydrodynamics at significantly higher temperatures (~100 K) based on first-principles calculations. In particular, by solving the Boltzmann equation for phonon transport in graphite ribbons, we predict that phonon Poiseuille flow and Knudsen minimum can be experimentally observed above liquid nitrogen temperature. Further, we reveal the microscopic origin of these intriguing phenomena in terms of the dependence of the effective boundary scattering rate on momentum-conserving phonon-phonon scattering processes and the collective motion of phonons. The significant hydrodynamic nature of phonon transport in graphite is attributed to its strong intralayer $sp^2$ hybrid bonding and weak van der Waals interlayer interactions. More intriguingly, the reflection symmetry associated with a single graphene layer is broken in graphite, which opens up more momentum-conserving phonon-phonon scattering channels and results in stronger hydrodynamic features in graphite than graphene. As a boundary-sensitive transport regime, phonon hydrodynamics opens up new possibilities for thermal management and energy conversion.

# 1 Introduction

Phonons are the dominant heat carriers in most dielectric materials. Phonon transport is usually diffusive and follows Fourier's law of heat conduction, which originates from momentum-destroying phonon scattering processes (R-scattering) such as Umklapp scattering (U-scattering), isotope scattering and impurity scattering. However, collisions between phonons are not necessarily momentum-destroying. Phonon-phonon normal scattering (N-scattering) indeed conserves the total momentum of phonons. If normal scattering is dominant, phonons can develop a nonzero drift velocity when subjected to a temperature gradient, very much like the viscous flow of a fluid driven by a pressure gradient. In analogy, such a heat conduction regime is called hydrodynamic phonon transport.[1] In general, at sufficiently small length scales and low temperatures, Fourier's law breaks down while hydrodynamic[1] and ballistic[2] transport of phonons emerge. Hydrodynamic transport has been widely studied in diverse systems such as viscous electron flow,[3,4] cold atoms[5] and quark-gluon plasmas.[6] However, for phonons, compared to the extensive studies on its ballistic transport,[7–9] phonon hydrodynamics has drawn much less attention. For decades in the 20th century, it was believed that phonon hydrodynamics occurs only at very low temperatures (~1 K).[10–13] More recently, investigations into low-dimensional (1D and 2D) materials have predicted that phonon hydrodynamic transport can also occur at significantly higher temperatures and over a wide temperature range.[14–17] However, it remains a long-standing challenge to identify bulk materials in which phonon hydrodynamics can take place at relatively high temperatures.

In the hydrodynamic regime, the collective drift of phonons leads to quite a few intriguing phenomena, such as second sound and phonon Poiseuille flow.[14,18,19] The former refers to the wave motion of heat in a phonon gas, which is similar to the propagation of sound waves in ordinary matter; while the latter is analogous to Poiseuille flow of a viscous fluid in a pipe, and is characterized by a phonon thermal conductivity that increases superlinearly with channel width.[20] Further, due to the presence of hydrodynamic effect, a phonon Knudsen minimum is expected near the transition from ballistic to Poiseuille heat conduction. The concept of Knudsen minimum was first introduced in the kinetic transport of rarified gases.[21] It refers to the phenomenon that the normalized flow rate in a channel with respect to the channel width ($d$) experiences a minimum when $d$ becomes comparable to the mean free path ($\lambda$) of the fluid particles, i.e., when the Knudsen number $Kn = \lambda / d$ is around 1. Such a Knudsen minimum was first observed experimentally by Knudsen[21] in 1909 and explained numerically by Cercignani[22] in 1963 by solving the fluid Boltzmann transport equation (BTE). Recently, a similar transport minimum was predicted for viscous electron flow[23,24] and in dilute granular systems.[25,26] The observation of phonon Knudsen minimum was reported many decades ago for heat flow in liquid helium between 0.25 K and 0.7 K.[27] However, its existence in solids above cryogenic temperatures remains an open question.[28]

Hydrodynamic transport is especially sensitive to boundary conditions, and thus offers new opportunities for controlling heat flow via size effect, which is key to efficient thermal management of electronics and thermoelectric energy conversion.[29] Despite past progress in modeling thermal transport in nanostructures,[30–34] studies of various size effects on heat flow were mostly based on the relaxation time approximation (RTA), which considers all phonon-phonon interactions as momentum-destroying and therefore is not suitable for discussing hydrodynamic transport. Description of phonon Poiseuille flow requires proper inclusion of normal scattering processes and phonon drift. Although full BTE solutions with mode-dependent phonon-phonon interactions have been obtained for single crystals, incorporating the size effect imposed by the heat flow boundaries still presents a significant challenge.

In this work, we develop a first-principles framework to study phonon hydrodynamics based on mode-specific phonon-phonon scattering details. In particular, we identify graphite as a remarkable three-dimensional material for high-temperature phonon hydrodynamic transport based on a recursive solution to the phonon BTE. This generalizes a range of 2D materials with strong hydrodynamic characteristics to bulk van der Waals (vdW) materials. Combining the relaxation times obtained from first-principles simulations with the Callaway model,[35,36] we solve the linearized BTE for graphite ribbons of different widths. Here, the classical Fuchs-Sondheimer solution[37–39] for the thermal conductivity of a thin film is extended to include phonon drift. We unambiguously reveal the existence of Poiseuille heat flow and superlinear size-dependent thermal transport – a non-trivial signature inherent to the hydrodynamic regime – in graphite ribbons at temperatures up to 90 K. In contrast, Poiseuille heat flow has only been previously observed in liquid helium below 1 K.[40] Further, we predict that the phonon Knudsen minimum – an unusual phenomenon that marks the crossover from ballistic to hydrodynamic transport – can be observed up to 90 K. The microscopic origin of these intriguing phenomena is elaborated based on the interplay between phonon-phonon normal scattering and boundary scattering in a pseudo material.

## 2 Results and Analysis

### 2.1 Collective phonon drift

A defining signature of hydrodynamic phonon transport is the collective drift motion of phonons, which mathematically manifests itself in the phonon distribution function. As bosons, phonons follow the Bose-Einstein distribution at thermal equilibrium. However, in the presence of a temperature gradient and when N-scattering dominates over R-scattering, phonons would equilibrate towards a displaced Bose-Einstein distribution[41] which is written as:

$$f_d = \frac{1}{\exp\left[\frac{\hbar(\omega + \mathbf{q} \cdot \mathbf{u})}{k_B T}\right] - 1} \quad (1)$$

where $\hbar$, $\omega$, $\mathbf{q}$, $k_B$ and $T$ denote the reduced Planck constant, phonon frequency, phonon wavevector, the Boltzmann constant, and temperature, respectively. The drift velocity, $\mathbf{u}$, is constant for all phonon modes regardless of polarization and wavevector. Assuming a small temperature gradient and drift velocity, Eq. (1) can be linearized as:

$$f_d = f_0 - f_0(f_0 + 1)\frac{\hbar}{k_B T}\mathbf{q} \cdot \mathbf{u} \quad (2)$$

where $f_0$ is the Bose-Einstein distribution.

Here, using a first-principles calculation framework[42–45] implemented by the ShengBTE package,[46] the phonon distribution function in an infinitely large graphite crystal under a constant temperature gradient in the zigzag ($x$) direction is calculated under a steady temperature gradient in the $x$ direction (zigzag) is calculated. To intuitively demonstrate the existence of phonon drift, we define a normalized deviation of the distribution function[14] from the equilibrium Bose-Einstein distribution as $\overline{df} = (f - f_0)/f_0(f_0 + 1)$, where $f$ is the exact solution to the phonon BTE. Figure 1a-b presents a map of the normalized deviation for the bending acoustic (BA) phonons, where $\overline{df}$ appears linear with $\mathbf{q}$ along the temperature gradient direction ($q_x$) at 100 K but the linearity disappears at 300 K (see also Fig. S1). Moreover, we observe in Fig. 1c that, at 100 K, $\overline{df}$ is linearly proportional to $q_x$ actually with the same slope for the first three phonon branches, indicating according to Eq. (2) a common drift velocity shared by all three branches. Our computation suggests that these three phonon branches contribute more than 90% of the total thermal conductivity (Fig. S2). With the dominant phonon branches sharing a common drift velocity, we conclude that phonon transport in graphite is in the hydrodynamic regime even at 100 K. Note that the deviation from linearity for high frequency phonons (Fig. 1c) is exaggerated due to the small values of $f_0$. In addition, these phonons contribute negligibly to the total thermal conductivity as demonstrated in Fig. S3.

To quantify how close the calculated phonon distribution $f$ is to the ideal displaced distribution $f_d$, we follow Ref. 15 and define a phonon drifting component as:

$$\rho_d = \frac{\sum C \overline{df} q_x}{\sqrt{\sum C \overline{df}^2} \sqrt{\sum C q_x^2}} \quad (3)$$

where $C$ is the specific heat and the summation is over all phonon modes (same below unless specified). It is easy to see that $\rho_d = 1$ when $f = f_d$. We plot the drifting component for graphite at different temperatures in Fig. 1d, and reveal that more than 90% of the phonon distribution comes from the collective drift motion at temperatures below 100 K. At room temperature, the drift component reduces to about 40%.

As mentioned earlier, the establishment of a collective phonon drift requires that N-scattering processes dominate over R-scattering. Quantitatively, we plot the mode-specific scattering rates at 100 K and 300 K in Fig. 1e-f. A characteristic frequency, $k_B T / 2\pi \hbar$, is also marked, below which phonons can be readily activated at a given temperature. Based on these scattering details, we reveal two mechanisms responsible for the suppression of hydrodynamic phonon transport at high temperatures. First, we observe that the significant dominance of N-scattering over R-scattering rates is reduced at higher temperatures, due to a generally steeper temperature dependence of R-scattering. Moreover, higher temperatures allow phonons of higher frequencies to be populated and contribute to heat flow. However, these high frequency phonons tend to have larger R-scattering rates.

## 2.2 Phonon Poiseuille flow and Knudsen minimum

We have demonstrated that at 100 K heat transport in graphite is hydrodynamic in nature, as it is dominated by drifting phonons (Fig. 1d). To further explore the regime of hydrodynamic phonon transport, we consider a graphite ribbon instead of a bulk crystal. In analogy to fluid flow in a pipe, phonon Poiseuille flow and Knudsen minimum become possible with the addition of the flow boundaries.[14,20]

We begin by discussing some general features of heat flow in an infinitely long graphite ribbon (Fig. 2a). When the sample width is much larger than the R-scattering mean free path (MFP), the transport is in the diffusive regime, and a uniform heat flux profile develops across the ribbon width (Fig. 2b) because phonon diffusion is dominated by R-scattering throughout the sample. In comparison, ballistic transport occurs when the sample is much narrower than the N-scattering MFP, which is also characterized by a nearly uniform heat flux profile since there is no scattering within the medium. Hydrodynamic transport takes place when the sample width is larger than the N-scattering MFP and but smaller than the R-scattering MFP. Here a quadratic heat flux profile is expected and the phenomenon is thus called phonon Poiseuille flow, analogous to viscous fluid flow.[14,20] The resistance to phonon Poiseuille flow comes mainly from an interplay between diffuse boundary scattering and phonon normal scattering. Briefly, under a temperature gradient a collective phonon drift develops via N-scattering. Since R-scattering is too weak, phonons near the center of the ribbon rarely experiences momentum loss. However, at the ribbon boundaries phonons scatter diffusely and lose a considerable amount of momentum. This leads to a decrease of heat flux from the center to the boundary of the ribbon. Due to the collective phonon drift, the effective boundary scattering rate differs from the Casimir limit.[2,47] Such difference underlies the pronounced size dependence of phonon Poiseuille flow as detailed below.

To accurately calculate the thermal conductivity and heat flux profile in the zigzag graphite ribbon (Fig. 2a), we have incorporated the Callaway model into the phonon BTE. Detailed derivations can be found in Method 4.3 and Supplementary Note 2. As demonstrated in a previous study[15] and in Fig. S4, Callaway model can capture the thermal conductivity of a bulk crystal with good accuracy in the presence of collective phonon drift, since it properly separates the momentum-conserving N-scattering from the resistive scattering processes. Here we further extend this framework to take into account the existence of flow boundaries as elaborated in Supplementary Note 2. This allows us to investigate phonon hydrodynamic transport in the context of size effect. Our calculation shows that the heat flux profiles in graphite ribbons of different widths at 70 K exhibit clear transitions between the ballistic, hydrodynamic and diffusive regime, when N-scattering dominates over R-scattering (Fig. S5a). This is a direct consequence of the large phonon drift at low temperatures (Fig. 1d). In comparison, at 300 K the hydrodynamic regime is missing because R-scattering dominates the phonon transport (Fig. S5b).

We now explore how the thermal conductivity ($k$) of a graphite ribbon varies as a function of ribbon width ($d$) at different temperatures, as shown in Fig. 3a. A finite sample width in general suppresses heat transport, but the width dependence of thermal conductivity may vary with temperature. To characterize the various transport regimes, we define a dimensionless scaling ratio $\alpha$ as $\alpha = \partial ln(k)/\partial ln(d)$ (Fig. 3b), which suggests that around width $d$ the thermal conductivity scales as $k \propto d^{\alpha}$. And $\alpha = 0$ relation is typical for diffusive transport (size-independent), while $\alpha > 1$ implies a superlinear size-dependence. At low temperature, where phonon hydrodynamic transport plays a critical role, we find that the scaling ratio with respect to the width increases first before reducing to zero eventually in the diffusive regime (Fig. 3b). Intriguingly, at low temperatures (say 50 K) when drifting phonons dominate the transport, a superlinear ($\alpha > 1$) width dependence of thermal conductivity is clearly observed (Fig. 3b), which is a direct consequence of the phonon Poiseuille flow.[14,20]

Such superlinear scaling can be understood based on the kinetic transport theory, in which the thermal conductivity $k$ is expressed as:

$$k = \sum C v^2 \tau \qquad (4)$$

where $v$ is the phonon group velocity, and $\tau$ is the phonon lifetime due to resistive scattering processes which can be expressed using Matthiessen's rule:[48]

$$\frac{1}{\tau} = \frac{1}{\tau_U} + \frac{1}{\tau_I} + \frac{1}{\tau_B} \qquad (5)$$

where the subscripts $U$, $I$ and $B$ denote Umklapp scattering, isotope scattering and boundary scattering, respectively. In the Poiseuille hydrodynamic regime, phonon scattering at the boundaries ($\tau_B$) dominates, which is often described by the Casimir theory:[2]

$$\frac{1}{\tau_B} = \frac{2|v_y|}{d} \qquad (6)$$

where $d$ is the sample width and $v_y$ is the phonon group velocity perpendicular to the boundary. Although the Casimir theory indeed captures diffuse phonon scattering at the flow boundaries, it only leads to sublinear width dependence for the thermal conductivity. To explain the unusual superlinear width dependence, one has to re-think how boundary scatterings effectively destroy phonon momentum when there is collective drift motion. When explaining the temperature dependence of thermal conductivity in crystalline He$^4$,[49] Gurzhi[50,51] proposed that phonons in the hydrodynamic regime should experience an effective boundary scattering given by:

$$\frac{1}{\tau_B} = \left(\frac{2\lambda_N}{d}\right)^2 \frac{1}{\tau_N} \quad (7)$$

where $\lambda_N$ is the N-scattering mean free path ($\lambda_N = v\tau_N$). Equation (7) relates phonon boundary scattering to N-scattering. Further, the boundary scattering rate is proportional to the inverse square of the ribbon width, rather than the inverse width as in Eq. (6). Compared to the Casimir theory, Gurzhi essentially suggests that phonons experience progressively less scatterings from the boundary in the hydrodynamic regime, as the ribbon width increases. The microscopic mechanism for this will be discussed in detail later. We note that although the relaxation time given by Eq. (7) is only an estimation for the effective boundary scattering, it does correctly capture the qualitative feature of the hydrodynamic regime – a superlinear width dependence of the ribbon thermal conductivity which can be readily seen from Eq. (4).

Using the superlinear scaling as a signature for phonon Poiseuille flow, we have further mapped out the different transport regimes with respect to sample width and temperature (Fig. 3c). At 90 K, the phonon Poiseuille flow could be observed when the sample width is within 1.5 – 2.8 μm, and increasingly wider range is predicted at lower temperatures.

As the ribbon width keeps decreasing, phonon transport eventually transitions from the hydrodynamic to the ballistic regime. At the transition width, the phonon Knudsen minimum is expected to occur. As discussed earlier, phonon Knudsen minimum for heat flow has only been reported in liquid helium between 0.25 K and 0.7 K.[27] Here we predict that in graphite ribbons the Knudsen minimum in phonon transport can be observed at temperatures as high as 90 K. In analogy to the dimensionless flow rate in rarified gas flow, we define a dimensionless thermal conductivity as:[27]

$$\bar{k} = \frac{3k}{Cv_s d} \quad (8)$$

where $v_s$ is the sound velocity in graphite.[52] The non-dimensionalization in Eq. (8) follows a definition traditionally used to describe Knudsen minimum in rarefied gas flow,[14,20] and differs from the normalized thermal conductivity plotted in Fig. 3a, which is the ratio of the ribbon thermal conductivity to the bulk value. Further, considering the broad distribution of phonon MFP, we define an average phonon MFP associated with N-scattering and weighted by the mode-specific contribution to the total thermal conductivity as: [53]

$$\langle \lambda_N \rangle = \frac{\sum Cv\lambda_N}{\sum Cv} \quad (9)$$

This allows us to use an effective phonon Knudsen number (defined as $Kn = \langle\lambda_N\rangle/d$) to characterize the heat flow. The dimensionless thermal conductivity with respect to the inverse phonon Knudsen number at different temperatures is shown in Fig. 3d, with the conductivity minima highlighted by the red dots. The existence of such phonon Knudsen minima again highlights the significant hydrodynamic nature of phonon transport in graphite. Strikingly, the phonon Knudsen minimum can be observed at temperatures up to 90 K, under the assumption of diffuse boundary scattering. The highest temperature to observe Poiseuille flow is expected to vary with respect to edge roughness as discussed in the dilute granular system.[25] Smoother boundary conditions facilitate the observation of Poiseuille flow and Knudsen minimum for relatively rough surfaces.[25] Therefore, employing diffuse boundary condition represents a conservative estimation for the temperature limit in practical cases.

Another evidence of phonon Poiseuille flow is found in the temperature dependence of the ribbon thermal conductivity,[49–51] which can be written as $k = Cv\lambda$ based on kinetic theory. Since the group velocity $v$ is essentially temperature independent, the heat-capacity-normalized thermal conductivity, $k/C$, should have the same temperature dependence as the phonon MFP, $\lambda$. In the ballistic regime, $\lambda$ is limited by boundary scattering thus mostly temperature independent, leading to an almost constant $k/C$ with varying temperature. In the diffusive regime, $\lambda$ decreases with increasing temperature due to increased U-scattering rates. However, with Poiseuille phonon flow, $\lambda$ is expected to increase as temperature rises. This is because in the hydrodynamic regime, the thermal resistance mainly comes from boundary scattering. As temperature increases, normal scattering between phonons becomes more frequent, effectively making the boundary scattering smaller (see Eq. (7)). Therefore, an increase of $k/C$ with temperature serves as an indicator of hydrodynamic heat flow. As shown in Fig. 3e-f, a clear increase of $k/C$ with rising temperature is observed when the graphite ribbon width is about 3 μm, suggesting hydrodynamic thermal transport. When the ribbon width is 250 nm, $k/C$ is essentially constant (slightly decreasing) with temperature, indicative of the ballistic regime. For a ribbon width of 10 μm, $k/C$ first increases at low temperatures but starts decreasing at higher temperatures, demonstrating a transition from hydrodynamic to diffusive transport. All these observations are consistent with the regime map shown in Fig. 3c.

At last we note that real samples are hardly perfect. In order to investigate the robustness of phonon hydrodynamics with respect to sample quality, we looked into the effect of vacancies as an example. We treat vacancy scattering as mass disorder,[54] and find that collective phonon drift motion is destroyed when the vacancy concentration is about 0.01%, but can still be observed when the concentration is about 0.001% (Fig. 4a-b). The regime map of graphite with 0.001% and 0.002% vacancy defects are shown in Fig. 4c-d. One can see that, in the presence of defects, phonon Poiseuille flow can become insignificant at low temperatures (Fig. 4d). This is because, at very low temperatures, both N-scattering and U-scattering are reduced, while defects scattering becomes dominant. Because phonon scatterings by defects are not momentum conserving, phonon hydrodynamic flow disappears at sufficiently low temperatures.

## 2.3 Microscopic origin of Knudsen minimum

The existence of phonon Knudsen minimum has been elusive in theory, particularly in solid materials. Here, we present a microscopic picture for the origin of the phonon Knudsen minimum. For the sake of simplicity but without loss of generality, we use a pseudo material modeled with the constant relaxation time approximation (Method 4.4). Further, we neglect R-scattering while only considering N-scattering and boundary scattering. For N-scattering, we compare cases with a scattering rate of $10^{10}$ Hz or zero, with zero indicating the absence of N-scattering. Before delving into the computational results, it may be instructive to note that in fluid flow, all scattering events are momentum-conserving and there is intrinsically no mechanism equivalent to phonon-phonon U-scattering. This means that the phonon flow in our pseudo material is actually a faithful analog of fluid flow. The existence of U-scattering will make the thermal conductivity eventually saturate at large width limit, as shown in Fig. S6, which is an effect the fluid flow does not have.

In Figure 5a we plot the calculated thermal conductivity of the pseudo material ribbon as a function of ribbon width, while Figure 5b shows the corresponding dimensionless thermal conductivity. When N-scattering is absent, phonons propagate ballistically until being diffusely scattered at the boundary, which translates to an effective MFP linearly proportional to the sample width. By writing Eq. (4) in terms of phonon MFP as $k = \sum Cv\lambda$, one can see that in ballistic transport the sample thermal conductivity linearly varies with the sample width. According to Eq.

(8), this means that the dimensionless thermal conductivity remains constant with varying sample width. When N-scattering is added into the picture, however, the thermal conductivity can be suppressed or enhanced, depending on the sample width (Fig. 5a). For small sample width, N-scattering serves mainly to increase boundary scattering, which reduces the sample thermal conductivity at a given width and leads to a smaller dimensionless thermal conductivity than the ballistic case. While for large sample width, N-scattering actually weakens boundary scattering, as previously pointed out by Gurzhi,[50,51] thereby leading to a larger dimensionless thermal conductivity than ballistic transport. These two considerations imply that there must be a minimum dimensionless thermal conductivity, that is, the phonon Knudsen minimum.

To understand the microscopic mechanism of the effect of N-scattering on phonon transport, we compute and plot the contribution to the total heat flux by phonons moving in different directions at various sample widths (Fig. 5c-e). In particular, we note that the MFP that affects the total thermal conductivity according to $k = \sum Cv\lambda$ should be regarded as the momentum-loss MFP, which characterizes the average distance that phonons travel before their momenta are randomized. In Fig. 5c-e, $\theta$ is defined as the angle between the phonon propagating direction and the boundary normal direction, as indicated in Fig. 5f. Phonons propagating parallel to sample boundaries are denoted by $\theta = 90°$. When the sample width is much smaller than the dominant N-scattering MFP (Fig. 5c), only phonons propagating almost parallel to sample boundaries are significantly affected by N-scattering, while most other phonons are not affected. More intuitively as shown in Fig. 5f, for phonons starting from the center of the ribbon, if their propagating direction is within the green region, they are less affected by N-scattering because they reach the boundary before experiencing significant N-scattering (corresponding MFP denoted by the dashed circle). On the other hand, phonons propagating along the direction in the red region will be redirected by N-scattering. When they are redirected towards the boundary, the boundary scattering rate is effectively increased. In other words, their contribution to the sample thermal conductivity is suppressed due to N-scattering. Such suppression becomes more severe as the sample width becomes larger (Fig. 5d), because more phonons (as indicated by the larger red region) are now affected by N-scattering before they reach the boundary (Fig. 5g). This leads to the initial decrease of the dimensionless thermal conductivity with increasing sample width (Fig.5b). If we define the sample thermal conductivity without N-scattering as $k^*$, and the suppression ratio due to N-scattering as $s$, then the thermal conductivity with N-scattering can be expressed as $k = k^* / s$. As discussed earlier, $k^*$ is proportional to $d$ in the ballistic regime. Since $s$ increases with increasing $d$ based on the analysis of Fig. 5c-d, the thermal conductivity $k$ should have a sublinear width dependence.

However, when the sample width becomes much larger than the N-scattering MFP, N-scattering processes can dramatically enhance the total thermal conductivity. Microscopically, this can be understood qualitatively from the picture of random walk.[20,51] When the sample width is much larger than the N-scattering MFP, each phonon experiences many N-scattering events before reaching the boundary. Because N-scattering conserves phonon momentum, the effective momentum-loss MFP then corresponds to the summation of all the individual propagation paths that eventually lead the phonon to the sample boundary, since only at the boundary is the collective phonon drift momentum lost. During such a random walk, the velocity of a phonon is randomized at each N-scattering event. Consequently, the average time for phonons to traverse a certain distance then quadratically depends on the square of distance, leading to an effective relaxation time given by Gurzhi[50,51] as $(d/2\lambda_N)^2 \tau_N$. This is distinct from the ballistic regime, where the time needed for phonons to reach the boundary is linearly proportional to the sample size. Such a modified boundary scattering term underlies the superlinear width-dependent thermal conductivity, when sample is sufficiently wide and the transport is dominated by drifting phonons. Combined together, the existence of both a sublinear and a superlinear width-dependent thermal conductivity dictates the existence of a phonon Knudsen minimum.

**2.4 From 2D to 3D**

The noteworthy hydrodynamic feature in graphene has been attributed to its high Debye temperature, quadratic phonon band and diverging Grüneisen parameter.[14] Here, the Debye temperature can be interpreted as the maximum acoustic frequency and a high Debye temperature implies that, at typical temperatures of interest, most of the activated acoustic phonons are located near the Γ point, which tend to have high N-scattering rates. We found that all these three favorable features are preserved in graphite (Fig. 6a-b), which contribute to its significant hydrodynamic behavior. This may be understood if we notice that the weak van der Waals interlayer interactions may not severely distort the lattice dynamics of individual graphene layers. However, we emphasize that the interlayer interactions do induce intriguing and unique features in graphite. First, the bending mode becomes stiffened in graphite (Fig. S7) and opens up a gap at the long-wavelength limit. More importantly, the interlayer interaction breaks the reflection symmetry carried by a single graphene sheet.[55] This renders many originally forbidden phonon-phonon scattering channels in graphene (such as **BA** + **BA** <=> **BA**, due to the vanishing coupling matrix[55]) accessible in graphite, which accounts for the lower thermal conductivity of graphite compared to graphene at room temperature. However, at low temperatures, as the thermally activated phonons mostly have small in-plane wavenumbers, the newly opened scattering channels are mainly N-scattering processes and actually facilitate the hydrodynamic transport. This is observed in Fig. 6c-d and Fig. S8, where comparisons between graphene and graphite are made in terms of the average linewidths (Fig. 6c), as well as the N- and R-scattering rates (Fig. S8), with the average linewidths defined as

$$\langle \Gamma_i \rangle = \frac{\sum C \left( 2\pi / \tau_i \right)}{\sum C} \quad (i = N, R) \tag{10}$$

As a result, the collective phonon drift is more significant and accounts for a larger percentage of the total thermal conductivity in graphite

## 3  Concluding remarks

In conclusion, we predict that phonon hydrodynamic transport can occur in graphite at up to 100 K. As discussed in ref. 14, the significant hydrodynamic feature in graphene is attributed to its high Debye temperature, quadratic phonon band and diverging Grüneisen parameter. Here we predict that all the three favorable features are preserved in graphite due to the weak van der Waals interlayer interactions. More importantly, we demonstrate that the hydrodynamic nature of phonon transport in graphite can be even more pronounced than in graphene. This is because the interlayer interaction, although weak, breaks a selection rule on phonon-phonon scattering and creates more N-scattering channels, which facilitates hydrodynamic transport. Thus we extend the strong hydrodynamic transport in 2D materials[15] to its vdW layered family. By taking into consideration of phonon drift motion, we extend the classical size effect into the hydrodynamic regime. We found that though N-scattering does not induce resistance itself, it can enhance or reduce the thermal conductivity by varying the effective boundary scattering rate. By solving the BTE for graphite ribbon, a superlinear size dependent thermal conductivity is predicted, which is a direct evidence of the phonon Poiseuille flow. Phonon Knudsen minimum in solids is predicted using a first principle calculation for the first time and a microscopic explanation is provided. Our results will hopefully stimulate further work into discovering novel material systems with significant phonon hydrodynamics, as well as new research into understanding and manipulating the phonon transport in the hydrodynamic regime.

## 4 Methods

### 4.1 Computational details

All the first-principle calculations are performed by Vienna Ab Initio Package (VASP)[56–58] with projector-augmented-wave (PAW) pseudopotentials and local density approximation (LDA) for exchange-correlation energy functional. To include the nonlocal vdW interactions, we use an explicit nonlocal functional of density named optB88 functional[59,60] as demonstrated in Ref. 61. The geometry optimization of the unit cell was performed with a 24×24×10 Monkhorst-Pack grid of **k**-point sampling. The second-order (third-order) force constants was calculated using a real space supercell approach with a 5×5×2 (4×4×2) supercell and 6×6×6 (8×8×6) k-grid. The Phonopy package[62] was used to obtain the second-order force constants. The thirdorder.py and ShengBTE packages[46] were used to obtain the third-order force constants and the exact solution to BTE respectively. A cutoff radius of about 0.4 nm was used, which corresponds to including the 5th nearest neighbor when determining the third-order force constants. To get the equilibrium distribution function and scattering rates using the third-order force constants, the first Brillouin zone was sampled with 50×50×8 mesh. Unless specified, the isotope content is 0.1% $^{13}$C.

### 4.2 Boltzmann transport equation with Callaway model

The static steady Boltzmann transport equation (BTE) with Callaway scattering's model[35] could be written as:

$$v \cdot \nabla f = -\frac{f - f_0}{\tau_R} - \frac{f - f_d}{\tau_N} \quad (11)$$

where $f_d$ is the displaced distribution function defined in Eq. (1). Unlike the general understanding that a phonon-phonon normal-scattering event demands complete conservation of phonon momentum, we propose that, in the context of heat transport, a scattering event should be considered N-scattering as long as phonon momentum is conserved in the direction of heat flow, because it does not contribute to thermal resistance.

The drift velocity can be determined from the momentum conservation of N-scattering process[36]

$$\sum (f - f_d)\hbar \mathbf{q} = \mathbf{0} \tag{12}$$

As a hexagonal crystal, a temperature gradient applied on the high symmetry line $\nabla_\alpha T$ could induce a drift velocity in the same direction $u_\alpha$

Combining Eq. (11) and (12) we get the drift velocity:

$$\frac{u_\alpha}{\nabla_\alpha T} = \frac{1}{T} \frac{\sum \omega \tau v_\alpha q_\alpha f_0(1+f_0)}{\sum \tau(\tau_R)^{-1} q_\alpha q_\alpha f_0(1+f_0)} \tag{13}$$

where τ and τ$_R$ the total lifetime and resistive scattering lifetime respectively, calculated using the force constants obtained in the last method section.

Applying τ and τ$_R$ from the DFT calculation to the above expression results in up to 30% larger $u_a$ than that obtained from recursive solution to the BTE within the investigated temperature range. In order to more faithfully reproduce the exact solution of BTE and the correct magnitude of the drift velocity, we rescaled the N-scattering and R-scattering rates as:

$$\tau_R^* = \frac{u_\alpha}{u_\alpha^{Exa}} \tau_R \tag{14}$$

$$\frac{1}{\tau_N^*} = \frac{1}{\tau} - \frac{1}{\tau_R^*} \tag{15}$$

where $u_\alpha^{Exa}$ is the drift velocity obtained from the recursive solution to the BTE. We note that this rescaling only introduces changes in the relative magnitude of the N-scattering and R-scattering rates, without affecting the total scattering rates based on the Matthiessen's rule. Such correction is reasonable, as it can reproduce the similar thermal conductivity and drift velocity compared with the exact solution to the BTE as demonstrated in Fig. S4 and without this rescaling, the predicted temperature limit will be slightly higher.

### 4.3 Solution for graphite ribbons

All the result presented in the phonon Poiseuille flow and Knudsen minimum sections are obtained by solving the BTE for a ribbon, in which the sample is infinitely long and thick, but with a finite width under a uniform temperature gradient in the longitudinal direction. The temperature is assumed to be constant in the transverse plane. Diffusive boundary conditions are applied, in which the reflection angle of phonons are independent of incident angle. The details can be found in the Supplementary Note 2.

### 4.4 Thermal conductivity of the pseudo-material with constant relaxation time approximation.

To elaborate on the effect of collective drift motion of the size effect, we constructed a pseudo material with adjustable scattering rates. More specifically, we consider a material with Debye temperature of 500 K with a sound velocity 1 km/s. A sphere is assumed for the first Brillouin zone and a mesh grid $50 \times 50 \times 50$ is used. We vary the N-scattering: $\tau_N^{-1} = 0$ Hz and $\tau_N^{-1} = 10^{10}$ Hz for comparison. The thermal conductivity of this material at 100 K is calculated using the method described in Supplementary Note 2.

### Acknowledgments

This work is supported by ONR MURI under award number N00014-16-1-2436 through UT Austin.

## Author contributions

Z.D., and T.-H.L. carried out the ab initio calculations. Z.D., J.Z., and V.C. developed the solution to the thin film geometry. Z.D., J.Z., B.S., M.L. and G.C. interpreted the results and wrote the manuscript. G.C. supervised the project. All the authors edited the manuscript.

## Competing financial interests

The authors declare no competing financial interests.

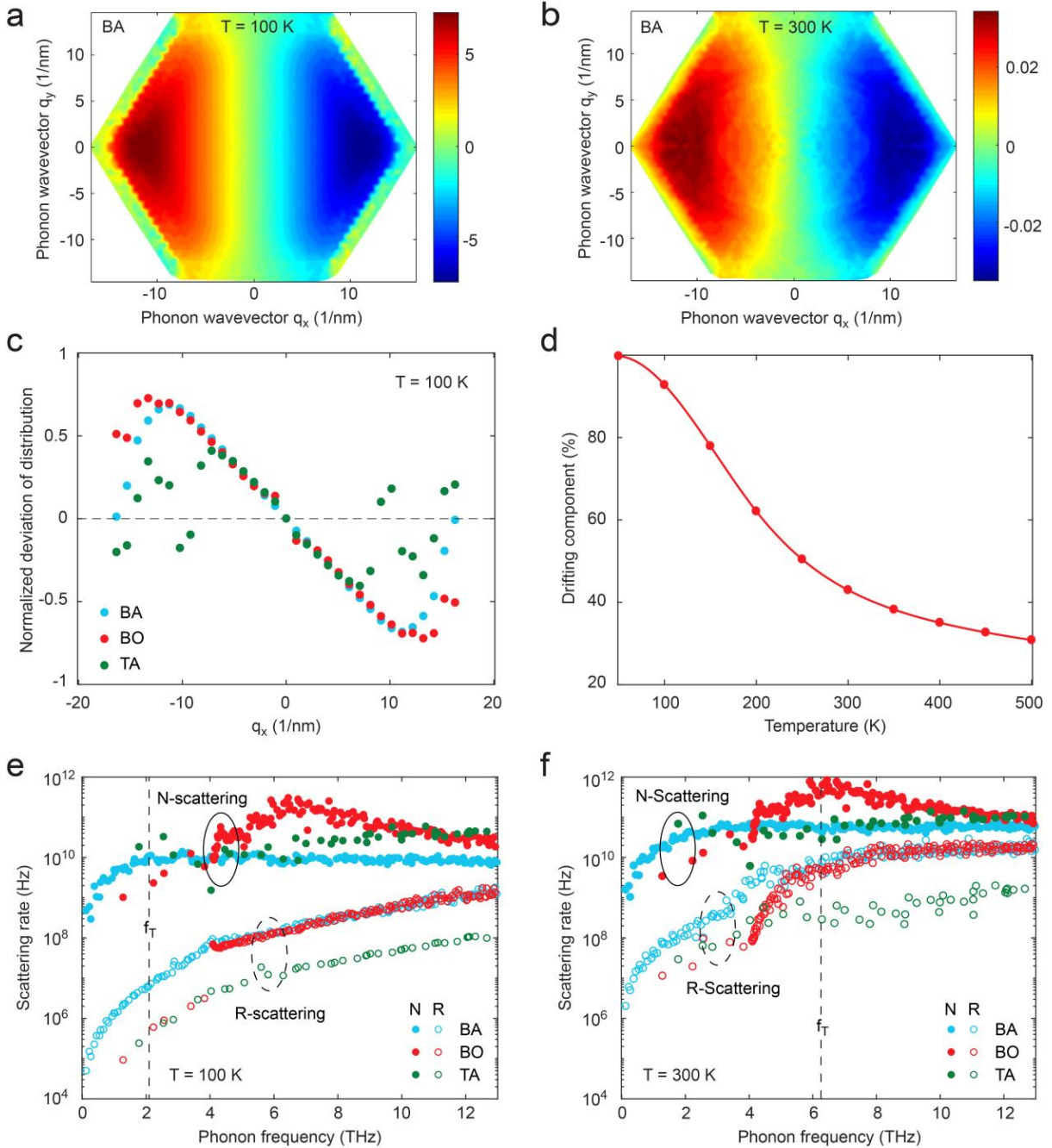

**Figure 1**. Signatures of phonon hydrodynamics in graphite (with 0.1% $^{13}$C) under a constant temperature gradient along the zigzag (*x*) direction. The normalized deviation of distribution function for the bending acoustic (BA) phonon mode in graphite at (a) 100 K and (b) 300 K. (c) The normalized deviation for the three lowest-frequency phonon branches (BO and TA stand for bending optical and transverse acoustic, respectively) along the *x* direction with $q_y = q_z = 0$ at 100 K. (d) Projection of the out-of-equilibrium phonon distribution onto the drifting distribution. Comparison of N-scattering and R-scattering rates with $q_z = 0$ at (e) 100 K and (f) 300 K, where a characteristic frequency $f_T = kT / 2\pi\hbar$ is marked.

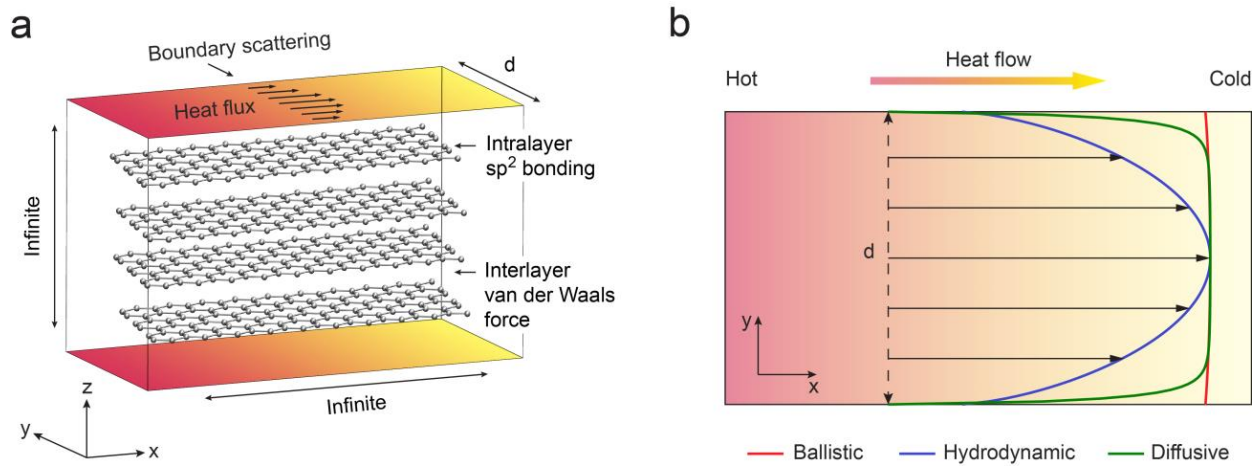

**Figure 2.** Schematic illustration of heat flow in a zigzag graphite ribbon. (a) Modeled graphite ribbon highlighting the finite width and phonon boundary scattering. (b) Heat flux profile in the ballistic, hydrodynamic and diffusive transport regime. The heat flux is essentially uniform in the diffusive and ballistic regimes, while a quadratic profile is expected in the hydrodynamic regime.

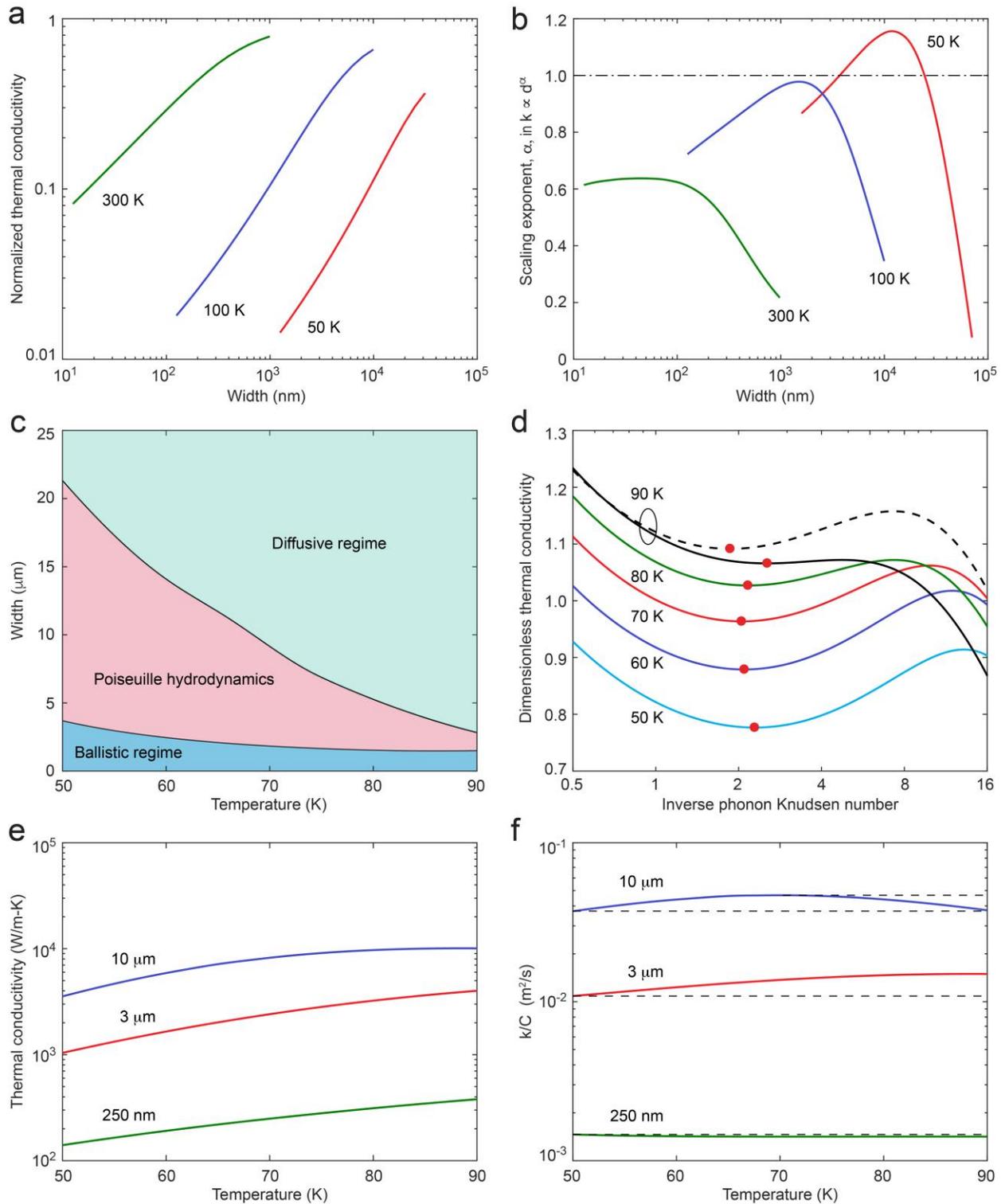

**Figure 3.** Thermal transport in graphite ribbons as a function of ribbon width and temperature. (a) Thermal conductivity variation with ribbon width at different temperatures. (b) Scaling of thermal conductivity with respect to ribbon width. A superlinear scaling window is observed at 50 K. (c) A map of various heat

transport regimes with respect to ribbon width and temperature, using the superlinear size-dependence as a signature of Poiseuille heat flow. (d) Variation of the dimensionless thermal conductivity with the inverse phonon Knudsen number at different temperatures. The solid lines are obtained for graphite with 0.1% $^{13}$C and the dashed lines are for isotopically pure graphite. Thermal conductivity as a function of temperature for graphite ribbons of different widths. (e) Thermal conductivity, and (f) thermal conductivity normalized by heat capacity, $k/C$. Increasing $k/C$ with rising temperature is an indicator of hydrodynamic Poiseuille heat flow. The horizontal dashed lines in (f) mark zero temperature dependence.

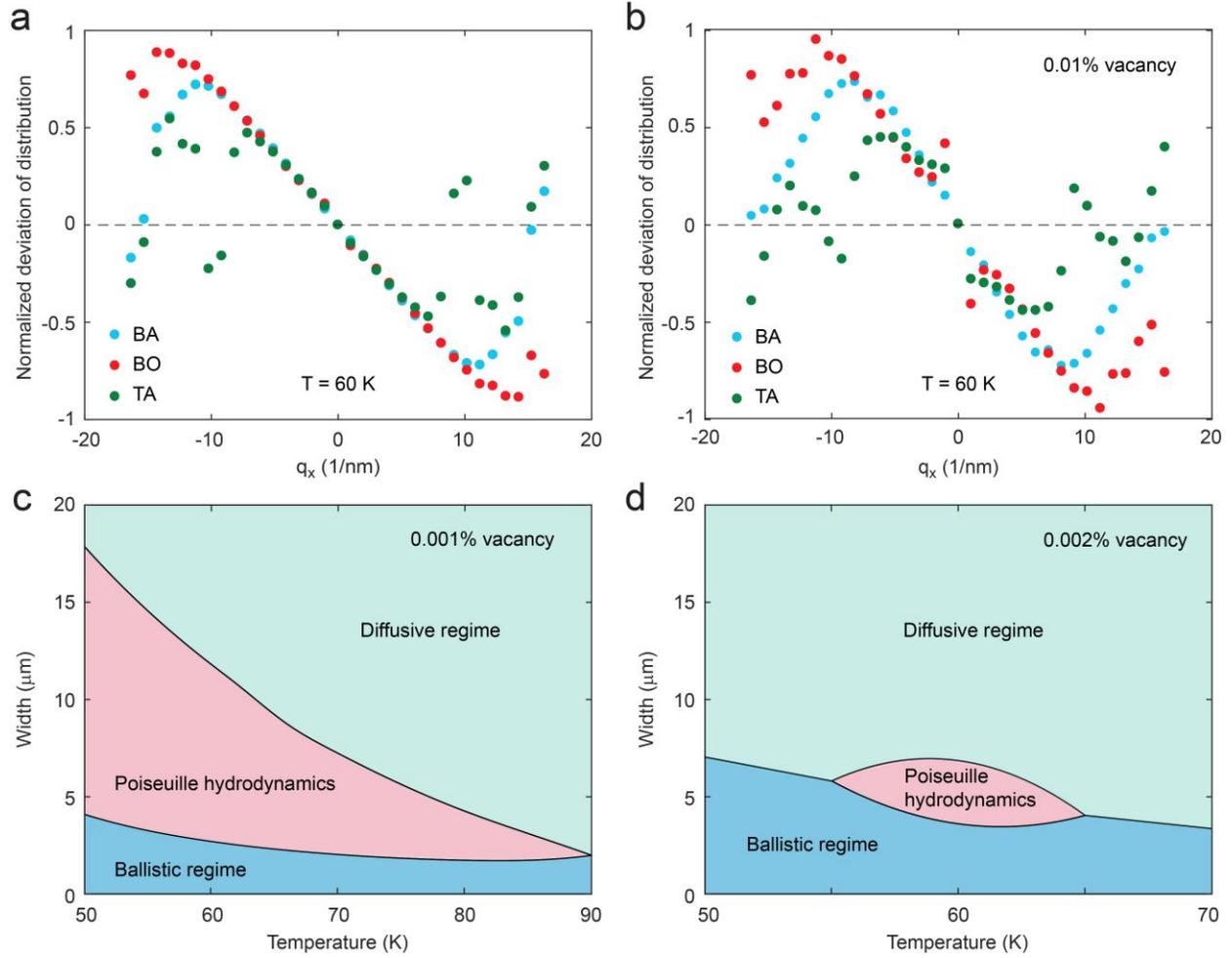

Figure 4. Thermal transport in graphite ribbons with defects. The normalized deviation of distribution function of the three lowest frequency phonon branches along the *x*-direction with $q_y = q_z = 0$ at 60 K in graphite with (a) 0.01% and (b) 0.001% vacancies. Maps of various heat transport regimes with respect to ribbon width and temperature, using the superlinear size-dependence as a signature of Poiseuille heat flow in graphite with (c) 0.001% (d) 0.002% vacancies

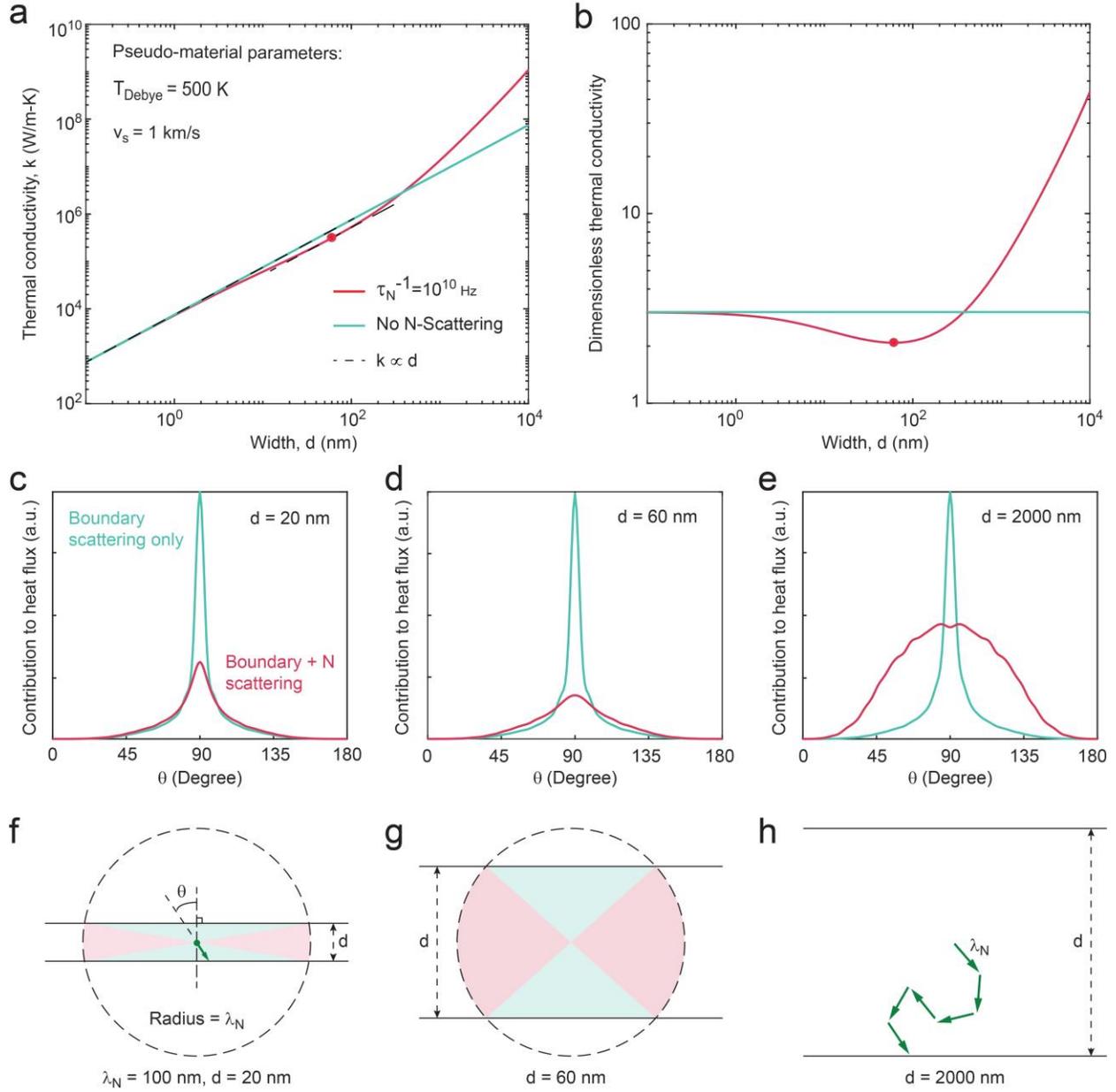

**Figure 5.** The microscopic origin of phonon Knudsen minimum. (a) Calculated thermal conductivity of a pseudo material as a function of sample width. (b) Variation of the dimensionless thermal conductivity with sample width. Phonon angle-dependent heat flux contribution with/without N-scattering for selected sample widths: (c) $d = 20$ nm, (d) $d = 60$ nm, (e) $d = 2000$ nm, assuming a constant N-scattering MFP ($\lambda_N$) of 100 nm. Illustration of the effect of N-scattering on the effective boundary scattering for (f) $d = 20$ nm (g) $d = 60$ nm. The dashed circle represents the N-scattering MFP. Boundary scattering of phonons traveling along a direction in the red region is affected by N-scattering, while phonons in the green region are not affected. (h) A schematic of the random walk of phonons in a very wide ($d = 2000$ nm) sample.

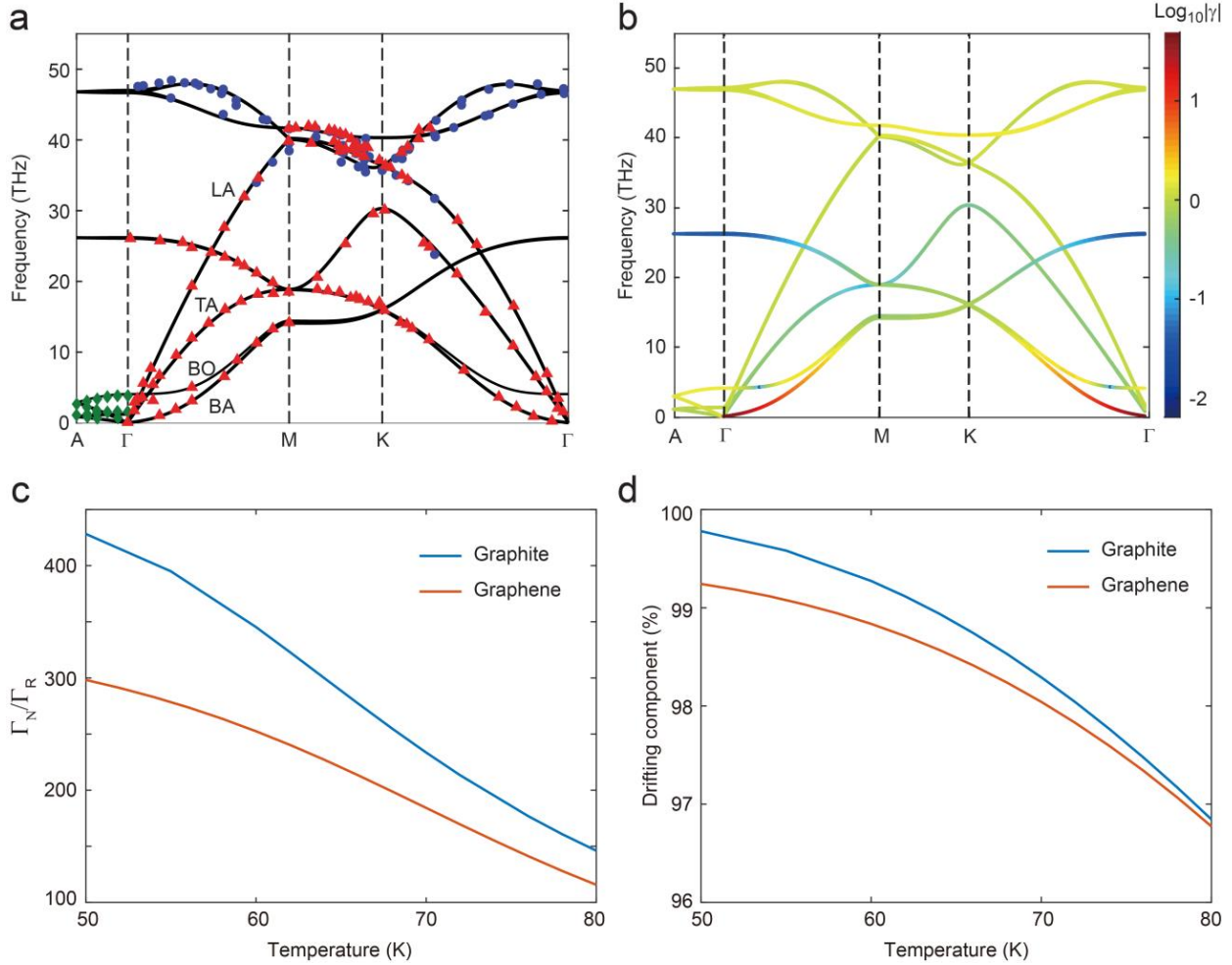

Figure 6. Comparison between graphite and graphene. Phonon band structure (a) and mode Grüneisen parameter (b) of graphite along the high symmetry lines. The blue circles are from Ref. 63, the red triangles are from Ref. 64, and the green squares are from Ref. 65. (c) Ratio of average N-scattering and R-scattering linewidth. (d) Projection of the out-of-equilibrium phonon distribution onto the drifting distribution.

# Table of Contents graphic

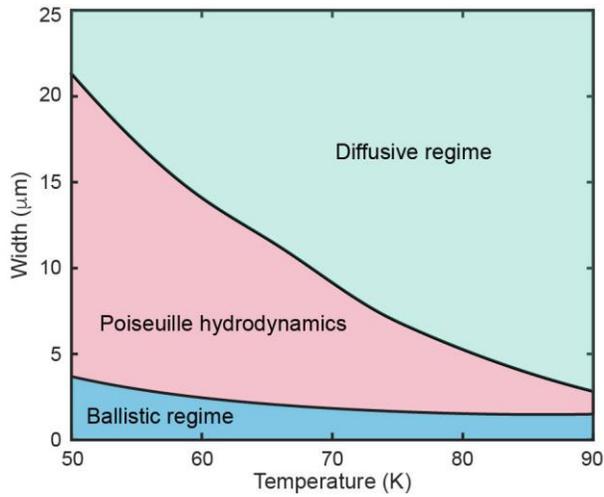 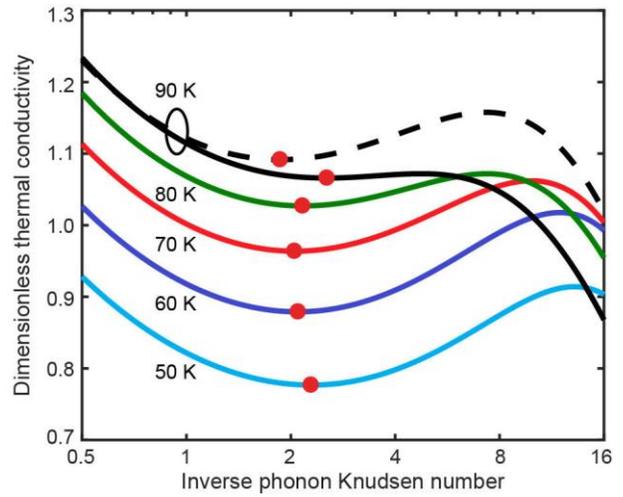